\documentclass[a4paper,11pt]{article}
\usepackage{pos}

\newcommand{\apj}{ApJ}
\newcommand{\apjs}{ApJS}
\newcommand{\apjl}{ApJL}
\newcommand{\aj}{AJ}
\newcommand{\aap}{A\&A}
\newcommand{\aaps}{A\&AS}

\title{Three in One: The VLBI Radio View of the X-ray Quasar RX J1456.0+5048}
\ShortTitle{Three in One: RX J1456.0+5048 with VLBI}

\author*[a,b]{S\'andor Frey}
\author[c,d]{Ingrid Tar}
\author[a]{Krisztina Perger}

\affiliation[a]{Konkoly Observatory, ELKH Research Centre for Astronomy and Earth Sciences,\\
Konkoly Thege Mikl\'os \'ut 15-17, H-1121 Budapest, Hungary}

\affiliation[b]{Institute of Physics, ELTE
E\"otv\"os Lor\'and University,\\
P\'azm\'any P\'eter s\'et\'any 1/A, H-1117 Budapest,
Hungary}

\affiliation[c]{GB Solutions Ltd.,\\ Lechner \"Od\"on fasor 8, H-1095 Budapest, Hungary}

\affiliation[d]{Department of Astronomy, Institute of Geography and Earth Sciences, ELTE
E\"otv\"os Lor\'and University,\\
P\'azm\'any P\'eter s\'et\'any 1/A, H-1117 Budapest,
Hungary}

\emailAdd{frey.sandor@csfk.org}
\emailAdd{tar.ingrid@gmail.com}
\emailAdd{perger.krisztina@csfk.org}

\abstract{RX\,J1456.0+5048 is a prominent X-ray source detected by \textit{ROSAT}. There is $\sim 100$-mJy level radio emission associated with the X-ray source. However, interferometric observations with increasing angular resolution revealed that three distinct objects located within 2 arcmin are responsible for the measured total flux density. Whether these radio sources lining up in the sky are physically associated or just seen close to each other in projection is not immediately clear. In fact, incorrect cross-identification of the X-ray, optical and radio sources can already be found in the literature. Here we summarise the current knowledge about this intriguing group of objects, where two of the three sources show compact radio emission detected with very long baseline interferometry (VLBI). We present a VLBI image of one of them for the first time, based on archival European VLBI Network (EVN) data taken at 5~GHz.}

\FullConference{%
  *** European VLBI Network Mini-Symposium and Users' Meeting (EVN2021) ***\\
  *** 12-14 July, 2021 ***\\
  *** Online ***
}

\begin{document}
\maketitle

\section{Introduction}

The \textit{ROSAT} All-Sky Survey (RASS) conducted in 1990 and 1991 was the first of its kind with an imaging telescope sensitive in the soft X-ray band ($0.1-2.4$~keV). The bright source catalogue (1RXS) contained more than 18,000 entries, including nearly 3,000 sources uniquely identified with extragalactic objects \cite{1999A&A...349..389V}. However, the RASS angular resolution was only $1.8^\prime$ and the positional accuracy $0.5^\prime$. The majority of \textit{ROSAT} extragalactic X-ray sources could be matched with radio sources in single-dish surveys with comparable angular resolution \cite{1995A&AS..109..147B,1997A&A...323..739B}. Among these, RX\,J1456.0+5048 could be associated to a radio source with flux density $232\pm21$~mJy in the 5-GHz Green Bank survey \cite{1996ApJS..103..427G}.

\begin{figure}[!ht]
\centering
\includegraphics[width= 0.75\columnwidth]{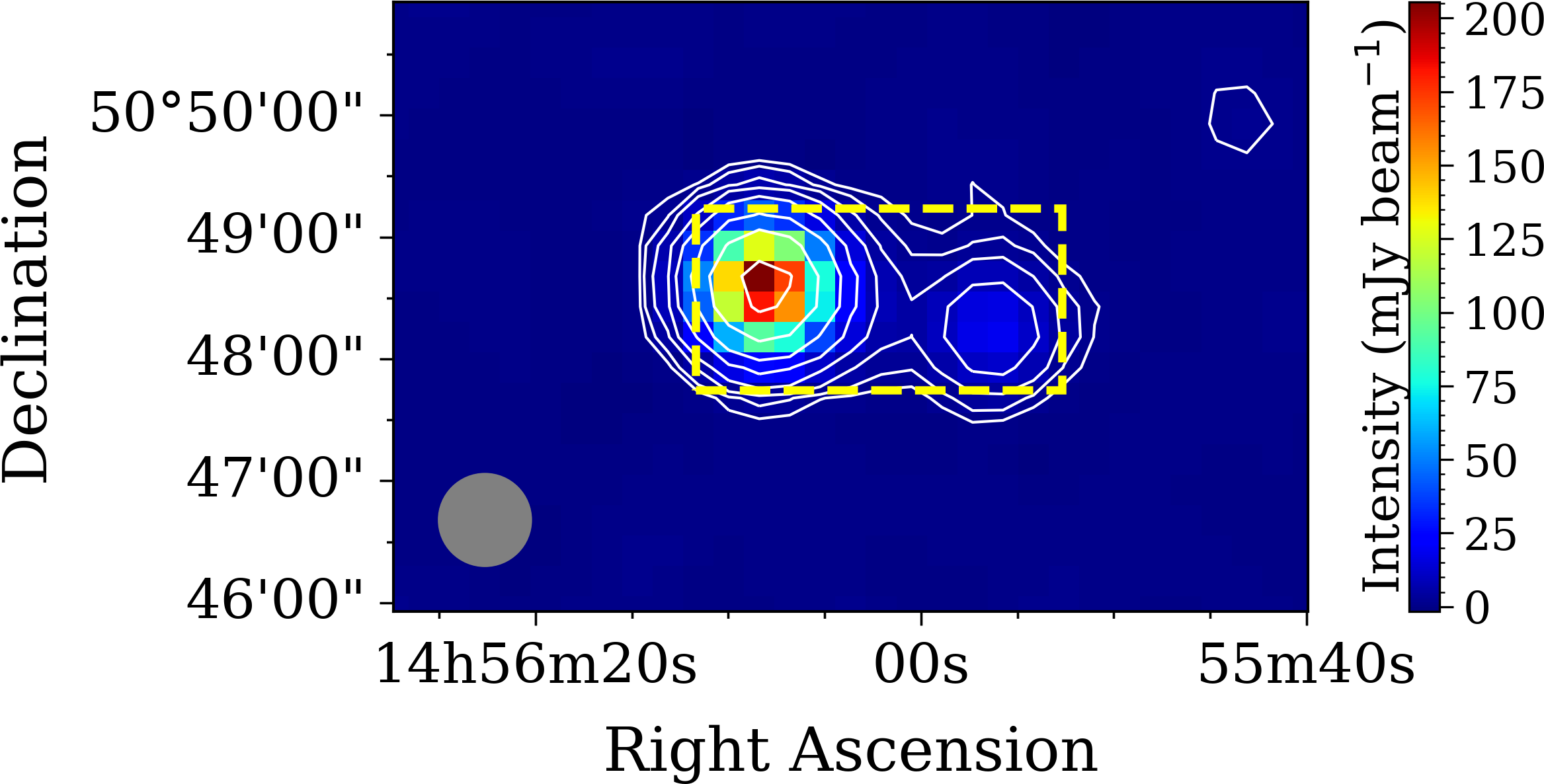}
\includegraphics[width= 0.88\columnwidth]{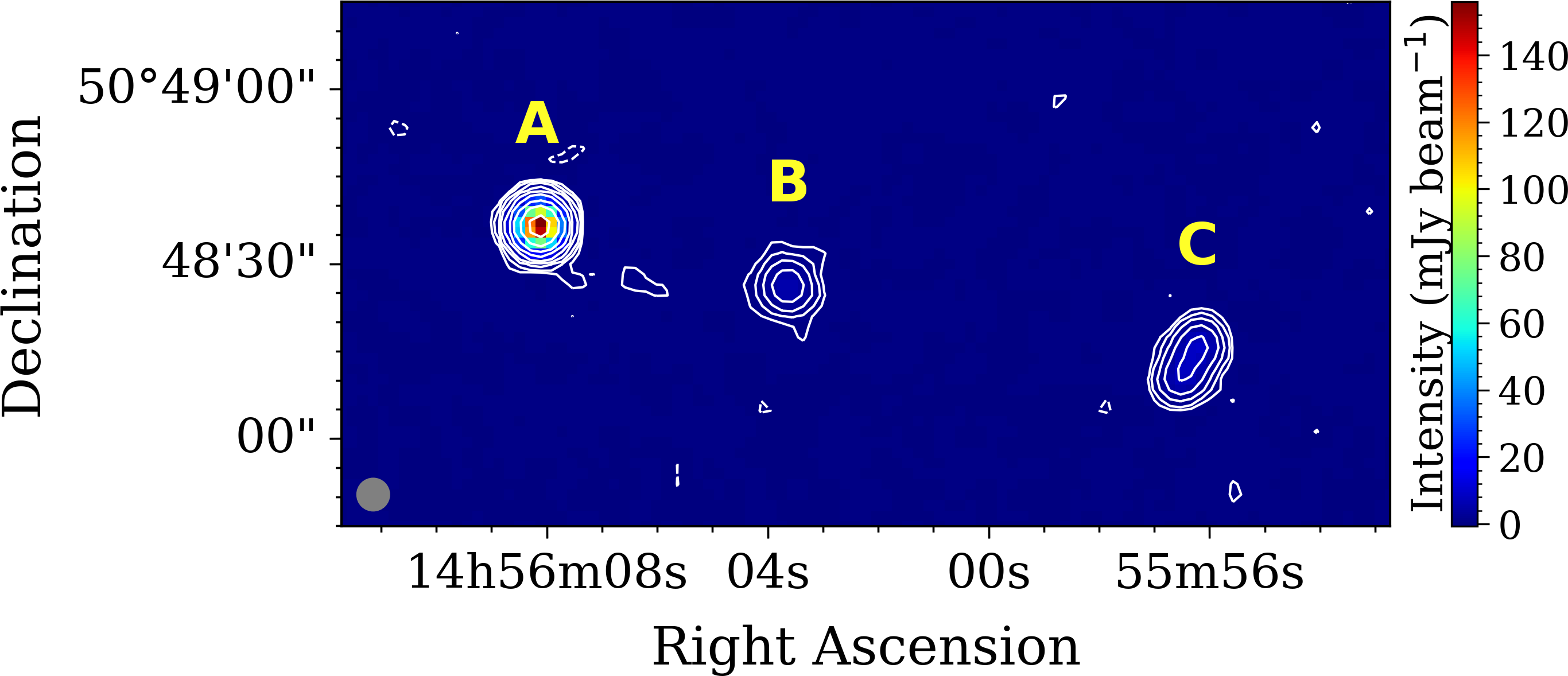}
\caption{1.4-GHz VLA image of the radio sources around the position of RX\,J1456.0+5048 from NVSS (top), and a higher-resolution view of the marked rectangular area in FIRST (bottom). The intensity colour scales are on the right-hand side, the half-power width of the Gaussian restoring beam is indicated in the bottom left corner of each image. In the FIRST image, the peak intensity is 155.9~mJy\,beam$^{-1}$, the lowest contours are at $\pm 3\sigma = 0.45$~mJy\,beam$^{-1}$, and the positive contours increase by a factor of 2. The labels A, B, and C mark the three distinct features discussed in the text.} 
\label{nvss-first}
\end{figure}

Radio interferometric observations with increasing angular resolution revealed that the source thought to be the radio counterpart of RX\,J1456.0+5048 is in fact composed of multiple features. Already in 1998, \cite{1998A&A...334..459B} noted that \textit{`the radio image of this [optically] faint object is somewhat confusing'}. Indeed, the 1.4-GHz Very Large Array (VLA) image from the NRAO VLA Sky Survey (NVSS, \cite{1998AJ....115.1693C}) shows two sources separated by $\sim 2^\prime$ in roughly east--west direction (Fig.~\ref{nvss-first}, top). Higher-resolution ($\sim 5^{\prime\prime}$) VLA observation at the same frequency from the VLA Faint Images of the Radio Sky at Twenty-centimeters (FIRST, \cite{1995ApJ...450..559B}) survey reveals that the brighter eastern feature in the NVSS image is further resolved into two sources (Fig.~\ref{nvss-first}, bottom). 

Obviously, in single-dish radio flux density measurements, as well as in lower-resolution interferometric images (like NVSS), some or all sources are blended together. The three sources, denoted by A, B, and C from east to west (Fig.~\ref{nvss-first}), form a nearly linear structure in the sky, raising the question whether they are physically associated components of the same radio source. Optical spectroscopic observations of one of them, source B, indicate redshift $z=0.48$ \cite{1998A&A...334..459B}. Assuming that all three components are at the same distance, this would correspond to $\sim700$~kpc projected liner size in the standard $\Lambda$CDM cosmology -- not unprecedented for a large radio galaxy. 
Unambiguous identification of active galactic nuclei (AGN) across the electromagnetic spectrum is essential for studying their broad-band spectral energy distribution (SED) and physical properties. Uncertain associations between radio, optical and X-ray sources should be avoided when constructing well-defined samples for statistical investigations of AGN classes. For example, \cite{1998A&A...334..459B} studied the evolutionary properties of a flux-limited sample of BL Lac objects selected from RASS. Another work analysed SEDs of more than 300 BL Lac objects based on extensive multi-frequency data \cite{2006A&A...445..441N}. RX\,J1456.0+5048 was included in both samples, but the validity of the source identification is questionable. 

Here we discuss high-resolution very long baseline interferometry (VLBI) data for two of the sources, A and B, and briefly mention some relevant multi-band information from the literature, to aid the correct classification of these objects located within $2^\prime$ in the sky.  

\section{VLBI observations}

\begin{figure}[!ht]
\centering
\includegraphics[width= 0.478\columnwidth]{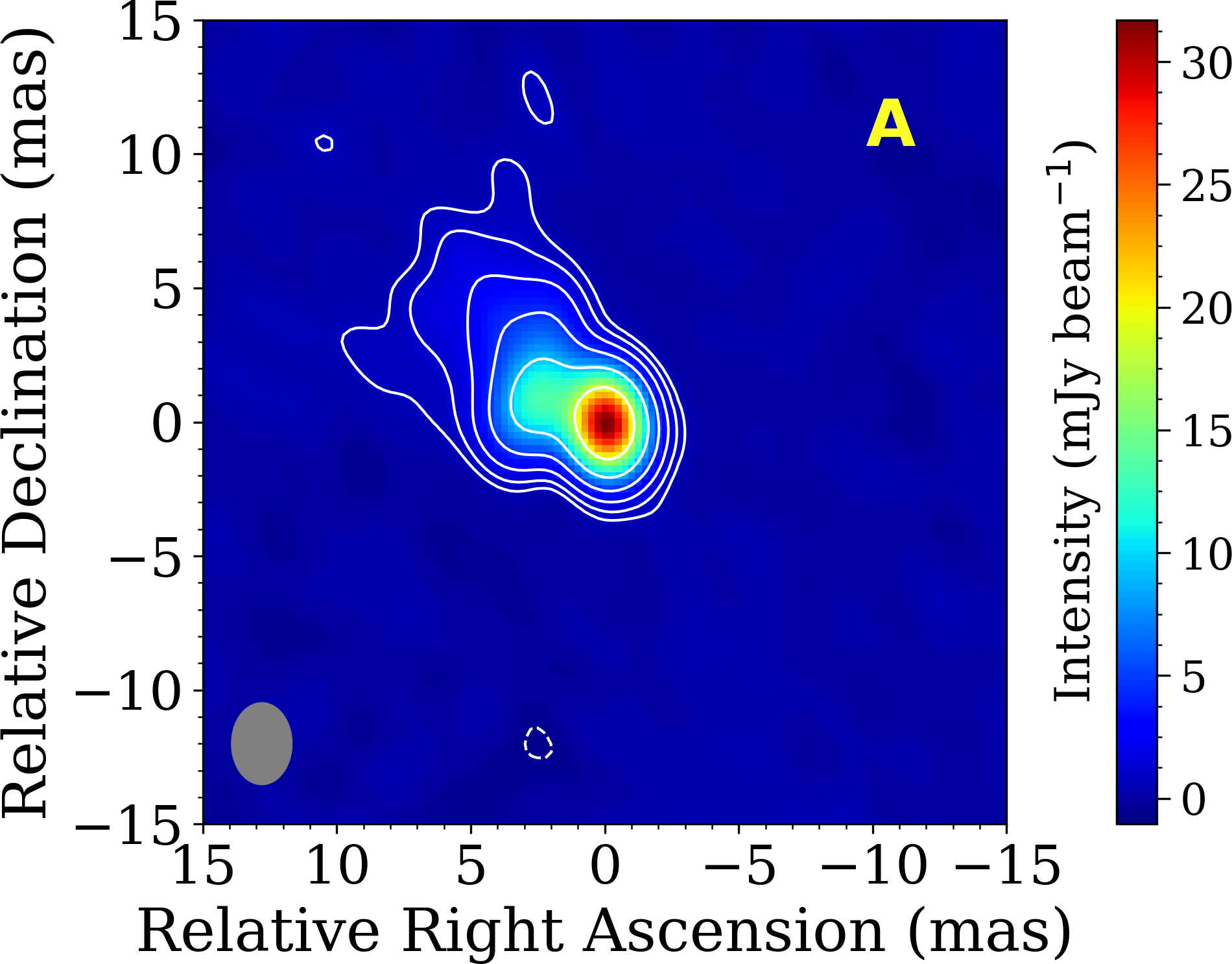}
\includegraphics[width= 0.49\columnwidth]{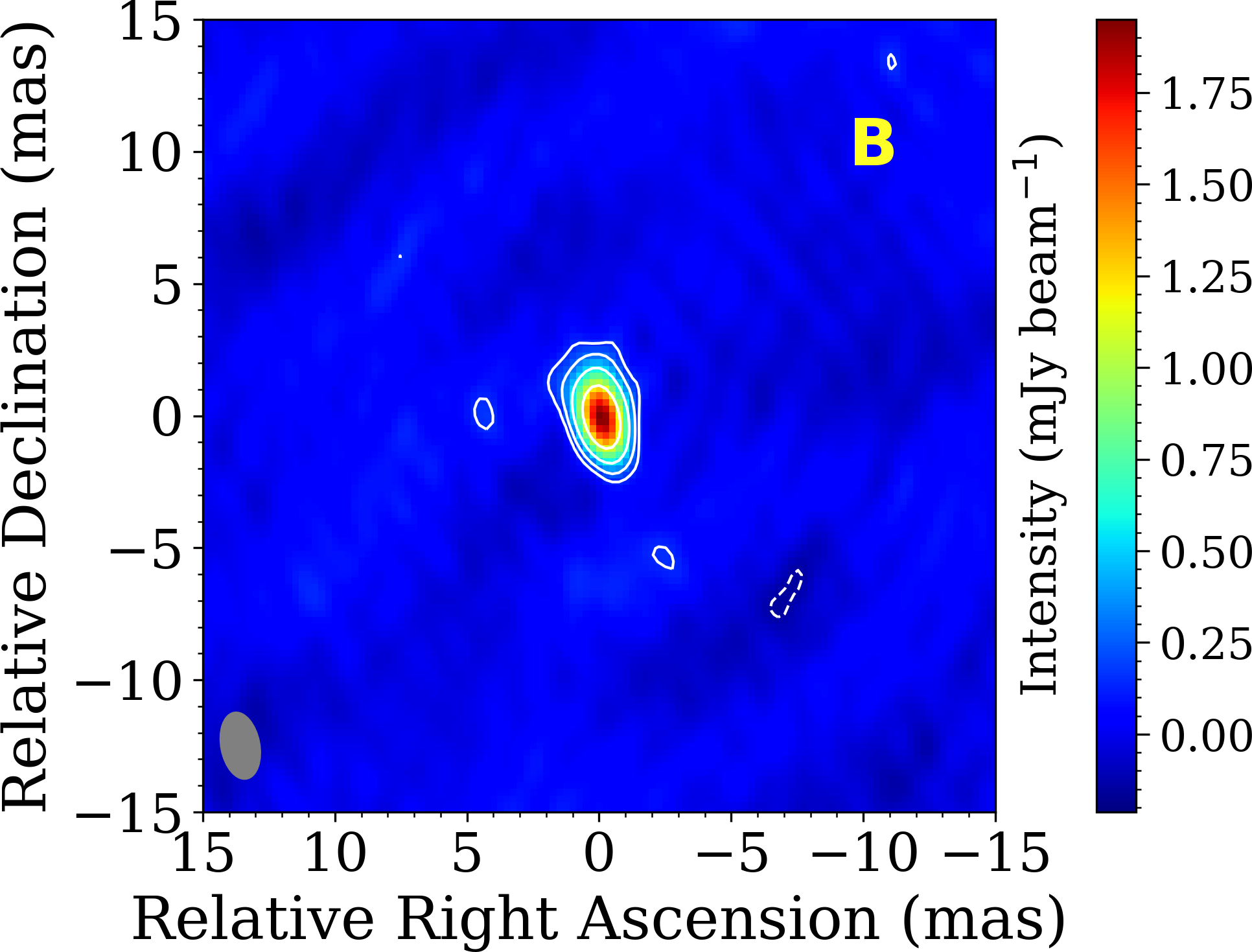}
\caption{5-GHz VLBI images of source A (left, taken with the VLBA on 2006 Aug 3, adopted from \cite{2007ApJ...658..203H}) and source B (right, taken with the EVN on 2015 Oct 28, project EW017A).} 
\label{vlbi}
\end{figure}

According to the FIRST data (Fig.~\ref{nvss-first}), the brightest of the three sources is A. Its 5-GHz Very Long Baseline Array (VLBA) image displayed in Fig.~\ref{vlbi} (left) is adopted from \cite{2007ApJ...658..203H}. The one-sided core--jet structure is typical for blazars. The precise absolute coordinates are right ascension $\alpha_{\mathrm A} = 14^\mathrm{h} 56^\mathrm{min} 08.119733^\mathrm{s}$ and declination $\delta_{\mathrm A} = +50^{\circ} 48^{\prime} 36.30050^{\prime\prime}$, with formal errors of $0.84$ and $0.53$ milliarcsec (mas), respectively \cite{2011AJ....142...89P}.

To make a 5-GHz image of source B (Fig.~\ref{vlbi}, right), we analysed publicly available data obtained from the European VLBI Network (EVN) archive (project EW017A, 2015 Oct 27--28, PI: Z. Wu). The participating radio telescopes were Effelsberg (Germany), Jodrell Bank Mk2 (UK), Westerbork (The Netherlands), Medicina, Noto (Italy), Onsala (Sweden), Sheshan, Nanshan (China), Toru\'n (Poland), and Yebes (Spain). This weak radio source was observed in phase-reference mode, with the calibrator J1439+4958 separated by $2.7^\circ$ from the target. The data were recorded in 32 spectral channels of 500 kHz width in each of the 8 intermediate frequency channels, in left and right circular polarizations, and correlated with 2~s averaging time at the EVN Data Processor. The total time spent on the target was about 2~h. The visibility data were calibrated in AIPS \cite{2003ASSL..285..109G} and imaged in Difmap \cite{1997ASPC..125...77S} in the standard way (see e.g. \cite{2020Symm...12..527K}). The source is well modelled with a weak (flux density $2.2 \pm 0.3$~mJy) compact (full width at half-maximum $0.6$~mas) circular Gaussian brightness distribution component. Its phase-referenced coordinates are $\alpha_{\mathrm B} = 14^\mathrm{h} 56^\mathrm{min} 03.63929^\mathrm{s}$ and $\delta_{\mathrm B} = +50^{\circ} 48^{\prime} 25.9759^{\prime\prime}$, with estimated errors within $1$~mas each.

\section{On the nature of the three sources}

\textit{Swift} X-ray Telescope observations with $18^{\prime\prime}$ resolution \cite{2013A&A...551A.142D} revealed that source B (1SWXRT J145603.6+504826) is \textit{the} X-ray source in the group. It is also the one which is the brightest in the optical (SDSS J145603.64+504825.9, \cite{2007ApJS..172..634A}), with spectroscopic redshift measured. On the other hand, object B is only a mJy-level compact source when observed with VLBI (Fig.~\ref{vlbi}, right). The $2.2$~mJy flux density of the mas-scale emission at 5~GHz is lower than that of the arcsec-scale emission at 1.4~GHz in FIRST, $9.45$~mJy \cite{1995ApJ...450..559B}. This could be attributed to a combination of a steep power-law radio spectrum, some extended structure resolved out with VLBI, and a flux density variability in time. 

It would be tempting to identify B with the core of a double-lobed radio AGN structure, where A and C are the nearly symmetric lobes. However, source A itself appears as a prominent jetted radio quasar with $\sim 60$~mJy flux density in its mas-scale compact structure, according to the VLBI data \cite{2007ApJ...658..203H} (Fig.~\ref{vlbi}, left). It is weak in X-rays and optical, with no redshift available. 

Based on \textit{Wide-field Infrared Survey Explorer (WISE)} data \cite{2012yCat.2311....0C}, the colours of both A and B position these objects in or very near the \textit{WISE} blazar strip \cite{2011ApJ...740L..48M}, suggesting that their infrared emission is dominated by non-thermal processes. 

There is comparably little observational information available about the westernmost source C, which appears resolved already on arcsec scale in the radio (Fig.~\ref{nvss-first}). 

In any case, sources A and B, both being blazar-like one way or another, should \textit{not} be confused with each other. Their markedly dissimilar radio, X-ray and optical properties and infrared colours imply that these are not gravitationally lensed images of the same background source. In the absence of redshift measurements for A and C, it is difficult to rule out a physical connection between any of the three sources. Most likely, the apparent proximity of the three objects is due to chance coincidence, and they are in different cosmological distances. However, a dual AGN system consisting of an X-ray-weak and radio-loud quasar (A), and an X-ray blazar with weak radio and optical emission (B), separated by $\sim 45^{\prime\prime}$ ($< 300$~kpc projected distance at $z=0.48$) is an intriguing possibility.

\acknowledgments
The EVN is a joint facility of independent European, African, Asian and North American radio astronomy institutes. Scientific results from data presented in this publication are derived from the following EVN project code: EW017. We thank the Hungarian National Research, Development and Innovation Office (OTKA K134213) for support.

\end{document}